\documentclass[ reprint
 amsmath,amssymb,
 aps, physrev,
]{revtex4-2}

\usepackage{graphicx}
\usepackage{dcolumn}
\usepackage{bm}
\usepackage{amsmath,amssymb}

\usepackage{braket}

\newcommand{\chem}[1]{{\mathrm{#1}}}
\newcommand{\ruo}{\chem{RuO_2}}
\newcommand{\ruA}{\chem{Ru}^{(A)}}
\newcommand{\ruB}{\chem{Ru}^{(B)}}
\newcommand{\ruX}{\chem{Ru}^{(X)}}
\newcommand{\ii}{\mathrm{i}}
\renewcommand{\Im}{\mathrm{Im}}
\renewcommand{\Re}{\mathrm{Re}}
\newcommand{\EF}{E_{\mathrm{F}}}
\newcommand{\eV}{\mathrm{eV}}
\newcommand{\odd}{\mathrm{odd}}
\newcommand{\even}{\mathrm{even}}

\newcommand{\soc}{\xi_{\mathrm{so}}}
\newcommand{\haihun}{\mathchar`-}

\makeatletter
\newcommand{\subscripts}[3]{%
  \@mathmeasure\z@\displaystyle{#2}%
  \global\setbox\@ne\vbox to\ht\z@{}\dp\@ne\dp\z@
  \setbox\tw@\box\@ne
  \@mathmeasure4\displaystyle{\copy\tw@_{#1}}%
  \@mathmeasure6\displaystyle{{#2}_{#3}}%
  \dimen@-\wd6 \advance\dimen@\wd4 \advance\dimen@\wd\z@
  \hbox to\dimen@{}\mathop{\kern-\dimen@\box4\box6}%
}
\makeatother

\begin{document}

\preprint{APS/123-QED}

\title{\textbf{
Neel vector dependent orbital Hall effect in altermagnetic RuO2
}}%

\author{Yuta Yahagi}
 \email{Contact author: yuta-yahagi@nec.com}
\affiliation{%
NEC Corporation, 
Business Development Department, Minato-ku, Japan}
\affiliation{
National Institute of Advanced Industrial Science and Technology, 
NEC-AIST Quantum Technology Cooperative Research Laboratory, Tsukuba, Japan
}%

\date{\today}

\begin{abstract}
Orbital Hall effect in the altermagnetic $\ruo$ is theoretically investigated. Transport calculations using the Kubo formula with a first-principles tight-binding model present a significant orbital Hall conductivity depending on the direction of the Neel vector. 
An pertuabative analysis explaines the mechanism by two roles of the spin-orbit interaction: converting the spin current into the parallelly polarized orbital current; recovering the orbital angular momentum anisotropically under the orthorhombic ligand field on the Ru sites.  
From an application perspective, this effect provides a promising method of generating orbital currents polarizing to arbitrary directions, realising an orbital source offering both controllability and efficiency.
\end{abstract}

\maketitle


\paragraph*{Introduction:}
Spin-torques induced by spin currents play vital roles in driving modern spintronics devices \cite{Manchon2019-ru, Ryu2020-ga, Song2021-zw, Shao2021-bw}. 
The magnetic random access memory (MRAM) is a representative application, in which the spin-Hall effect (SHE) \cite{Hirsch1999-zx,Sinova2015-sp} in the writing layer injects spin angular momentum to the free layer, exerting spin-orbit torque without damaging the barrier layer.
Gaining efficiency of spin torque generation is of great interest to reduce the operation current and improve device performances.

Recently, there has been an increasing number of studies on the "magnetic" SHE (MSHE) in magnetic materials due to its attractive features \cite{Miao2013-bg,Taniguchi2015-vn,Zelezny2017-vj,Kimata2019-ix,Davidson2020-uz, Yahagi2021-jh, Yahagi2022-dt}. 
Unlike to the conventional SHE in nonmagnetic materials, it can generate spin currents polarizing arbitrary direction as it is free from the restrictions imposed by time-reversal symmetry (TRS) \cite{Seemann2015-ny}, enabling the field-free switching of a perpendicularly magnetized film \cite{Baek2018-dn}.
In particular, a subset of anti-ferromagnets breaking TRS, classified as altermagnets \cite{Smejkal2022-za,Smejkal2022-yx}, have attracted considerable attention in this area since the discovery of a gigantic MSHE, so-called spin splitter effect, in the rutile $\ruo$ \cite{Gonzalez-Hernandez2021-ke, Karube2022-gb}.
Its origin does not rely on the spin-orbit interaction (SOI) but the exchange interaction coupled with the crystalline symmetry, sustaining such the larger effect than other relativistic phenomena. 
Moreover, the polarization of the spin current is along with the Neel vector direction. 
Therefore, the advantages of using $\ruo$ and the similar type of d-wave altermagnet are not only the large effect but also its controllable spin current, which is favorable for device application.

Another important trend in this field is the orbitronics, use of the orbital angular momentum (OAM) and the orbital current \cite{Go2021-bs,Jo2024-ng}. 
Although the OAM in a crystal is quenched in equilibrium due to a crystalline field or a ligand field, it is not under an external field; 
The OAM is recovered by the transition moment of the external field, leading orbital responses such as the orbital Hall effect (OHE) \cite{Tanaka2008-lf,Kontani2007-rv, Kontani2009-pn}.
The orbital current can also exert a spin torque, thus it has been gaining attention as a promising alternative or a cooperative phenomenon of the spin current.
Similarly to the SHE, the OHE in a nonmagnetic material is constrained by the TRS, whareas the magnetic OHE (MOHE) in a magnetic material is not. 
To date, the MOHE has been seldom studied except in simple ferromagnetic metals \cite{Salemi2022-eu,Salemi2022-rc}. 

While both altermagnetism and orbitronics are growing subjects, there is a surprising paucity of studies investigating the orbital transports in altermagnets.
In this work, we theoretically investigate the OHE in altermagnets through a first-principles study on the $\ruo$ as a representative material.
We also elucidate its microscopic mechanism qualitatively by using an analytical model based on a perturbation theory.

\paragraph*{Theory:}
Within the linear response theory based on the Kubo formula \cite{Kubo1957-jy}, the orbital Hall conductivity (OHC) is formulated as a current--orbital current correlation function,
\begin{equation}\label{eq:Kubo-formula1}
\sigma^{L_\mu}_{yx}=\lim_{\eta\to +0}\frac{\ii}{\hbar \Omega} \int_{0}^{\infty}dt \int_{0}^{\beta}d\lambda
\braket{\hat{J}_x(-\ii \hbar \lambda) \hat{J}_y^{L_\mu}(t)} e^{-\ii \omega t-\eta t},
\end{equation}
where $\hat{J}_\alpha$ is the $\alpha$-component of the current operator and $\hat{J}_\alpha^{L_\mu}$ is that of the orbital current operator polarized to $\mu$.
Here, the later is defined as 
$\hat{J}_\alpha^{L_\mu} \equiv -\frac{\hbar}{4e}( \hat{J}_\alpha \hat{L}_\mu+\hat{L}_\mu \hat{J}_\alpha ),$ 
with the intra-atomic orbital operator $\hat{L}_\mu$.
$\braket{\hat{X}}$ denotes the thermo-statistical average.
$\Omega$ is the volume. $\beta$ is the inverse temperature.
The spin Hall conductivity (SHC) can also be obtained from Eq. \eqref{eq:Kubo-formula1} by replacing $\hat{L}_\mu$ to the spin operator, $\hat{S}_\mu$. 

Hereinafter, we consider the DC transport ($\omega\to +0$) at the zero temperature ($\beta\to +\infty$), and assume a constant life-time approximation, resulting
\begin{align}
\label{eq:Kubo-formula2}
    \sigma^{L_\mu}_{yx}&=\sigma^{L_\mu, \odd}_{yx}+\sigma^{L_\mu, \even}_{yx},\\
\label{eq:Greenwood-like}
    \sigma^{L_\mu, \odd}_{yx} &= \frac{\hbar}{\pi\Omega} \Re \sum_{\bm{k},n,m}
        \braket{n\bm{k}|\hat{J}^{L_\mu}_y|m\bm{k}} \braket{m\bm{k}|\hat{J}_x|n\bm{k}} \rho_{n\bm{k}}\rho_{m\bm{k}}, \\
\label{eq:TKNN-like}
    \sigma^{L_\mu, \even}_{yx} &= \frac{2\hbar}{\pi\Omega} \Im \sum_{\bm{k},n\neq m} \frac{\braket{n\bm{k}|\hat{J}^{L_\mu}_y|m\bm{k}} \braket{m\bm{k}|\hat{J}_x|n\bm{k}}}{(E_{n\bm{k}}-E_{m\bm{k}})^2}
    f(E_{n\bm{k}})\left \{ 1-f(E_{m\bm{k}}) \right \},
\end{align}
where $E_{n\bm{k}}$ and $\ket{n\bm{k}}$ are the dispersion of band $n$ and corresponding eigen state.
$\rho_{n\bm{k}}$ is the spectral weight at the Fermi energy $\EF$. $f(E_{n\bm{k}})$ is the Fermi--Dirac distribution function.
The first term, $\sigma^{L_\mu, \odd}_{yx}$, is the time reversal odd (T-odd) term describing dissipative transports, and the second term, $\sigma^{L_\mu, \even}_{yx}$, is the time reversal even (T-even) term associated with the orbital Berry curvature. 

The $\ruo$ has the rutile structure with $a=b\simeq 4.5$ and $c\simeq 3.1$ in Angstrom \cite{Mehtougui2012-ag}.
In the anti-ferromagnetic phase, it has two $\chem{Ru}$ sites, $\ruA$ and $\ruB$, with opposite magnetic moments each other. 
Each of them is surrounded by six $\chem{O}$ forming a distorted octahedron whose local axis alternates with respect to the magnetic sub-lattice, originating distinctive features of altermagnetism.
The Neel vector $\bm{n}$ strongly prefers the $c$-axis ($[001]$) due to a large axial magnetic anisotropy.
In this study, on the other hand, we consider not only $\bm{n}\parallel[001]$ but also $\bm{n}\parallel[100]$ for comparison.

To obtain the Hamiltonian of the $\ruo$ from first-principles, we conduct an electronic structure calculation based on the density functional theory (DFT), then construct a Wannier tight-binding (TB) model. 
We use a full-potential local orbital basis method implemented in the FPLO code \cite{Koepernik1999-fr,Opahle1999-uh}. 
Following previous works, we employ the generalized gradient approximation (PBE96) functional \cite{Perdew1996-ki} with a Hubberd-U ($U=3\ \mathrm{eV}$) for the $\chem{Ru}-4d$ orbitals that stabilizes the magnetic ground state \cite{Gonzalez-Hernandez2021-ke}.
The SOI is considered within a fully-relativistic scheme of the four-component Kohn--Sham--Dirac equation. 
The Wannier TB model is directly obtained from a FPLO Hamiltonian by imposing an energy threshold of $0.001\ \eV$ into its matrix elements.

\begin{figure}[h]
    \centering
    \includegraphics[width=0.5\linewidth]{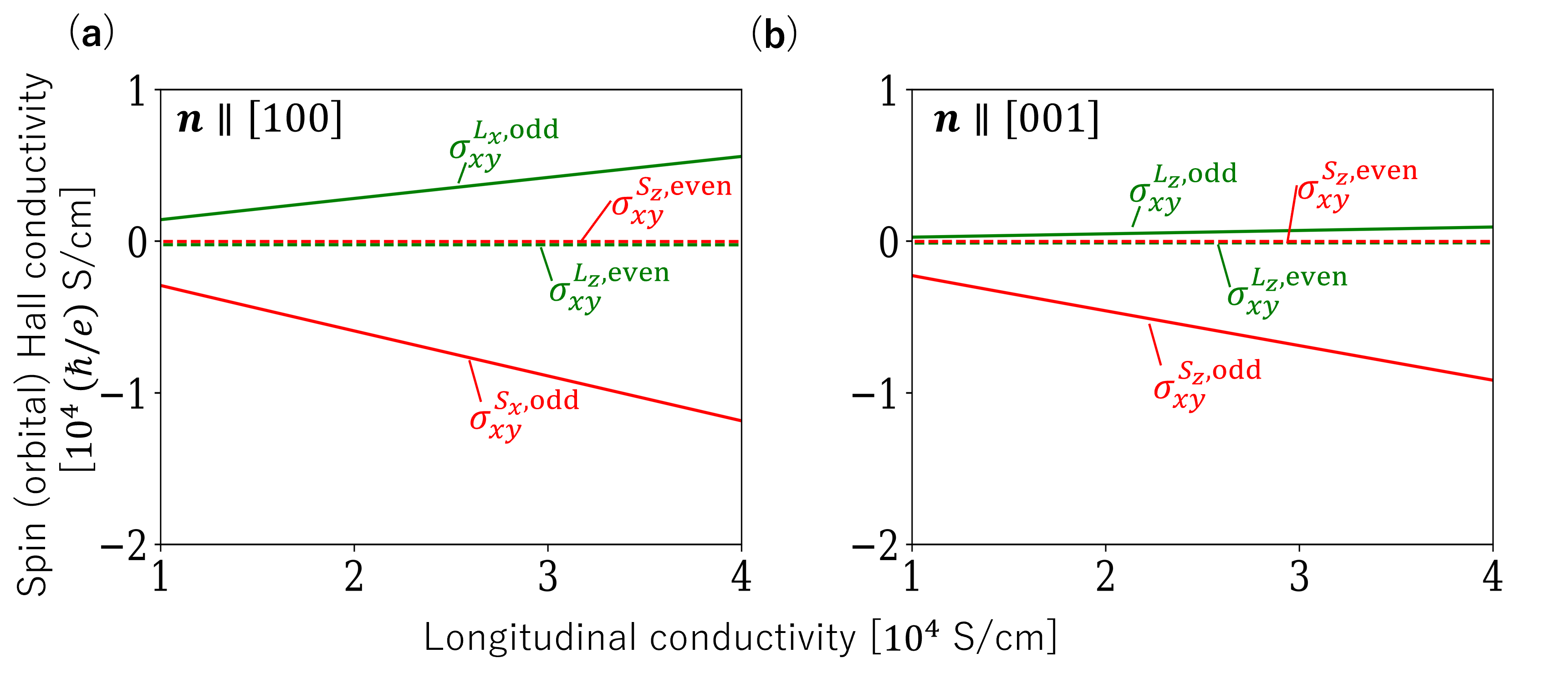}
    \caption{(color online) Orbital (spin) Hall conductivity $\sigma^{L_\mu}_{yx}$ ($\sigma^{S_\mu}_{yx}$) as a function of the longitudinal conductivity when the Neel vector is (a) parallel to the $a$-axis ($\bm{n}\parallel[100]$) or (b) parallel to the $c$-axis ( $\bm{n}\parallel[001]$). T-even and T-odd terms are displayed separately.}
    \label{fig:OHCvsCond}
\end{figure}

\paragraph*{Computational results:}
The computational results of the OHC (SHC) with different $\bm{n}$ are shown in Figs. \ref{fig:OHCvsCond}. 
The orbital or spin polarization is selected so that parallel to $\bm{n}$. 
Overall, the T-even terms are negligibly small and both the OHE and the SHE are dominated by T-odd terms. 
What stands out in the Fig. \ref{fig:OHCvsCond}(a) is the presense of not only a huge SHC consistent with the previously-reported spin splitter effect \cite{Gonzalez-Hernandez2021-ke} but also a sufficiently large OHC with the same order of magnitude.
It is approximately 5 times larger value than the MOHE in simple ferromagnetic metals.
Interestingly, comparing Fig. \ref{fig:OHCvsCond}(a) and Fig. \ref{fig:OHCvsCond}(b), the OHC shows strong dependency on the Neel vector as it drastically decreases by rotation from $\bm{n}\parallel[100]$ to $\bm{n}\parallel[001]$.

\begin{figure}[h]
    \centering
    \includegraphics[width=0.5\linewidth]{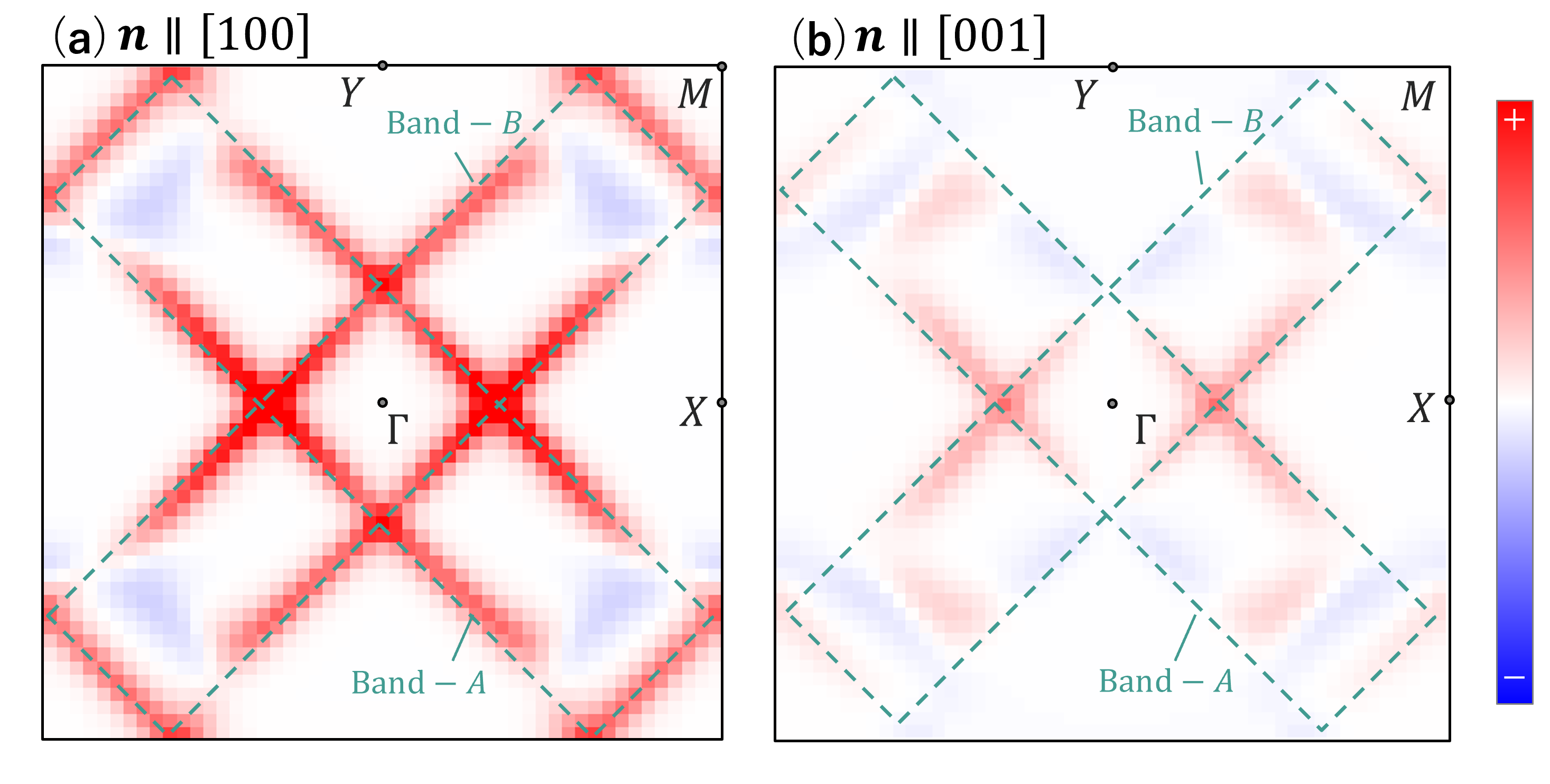}
    \caption{(color online) Momentum-resolved map of the orbital Hall conductivity across the first Brillouin zone at $k_z=0$ when the Neel vector is (a) parallel to the $a$-axis ($\bm{n}\parallel[100]$) or (b) parallel to the $c$-axis ( $\bm{n}\parallel[001]$). Two dominant bands mainly consisting with the 4d-orbitals of $\ruA$ or $\ruB$ are emphasized by dashed lines and labeled as band-$A$ and $B$, respectively.}
    \label{fig:k-resolved}
\end{figure}

Figs. \ref{fig:k-resolved} provide momentum-resolved intensity maps of the OHC contribution obtained by the integrand in the right-hand-side of Eq. \eqref{eq:Greenwood-like}.
These results indicate that the $\chem{Ru}-4d$ bands (band-$A$ and $B$) are main contributor of the OHC in both cases as emphasized in the dashed lines. 
Comparing Figs. \ref{fig:k-resolved} (a) and (b), we can see that there are significant differences in intensity depending on the direction of the Neel vector. 
The reason of such the difference is further analyzed.

\paragraph*{Perturbative analysis:}
To clarify a mechanism of the Neel vector dependence, we derive a simplified formula for the OHE in the $\ruo$.
Similar to the case of the MSHE or the longitudinal conductivity, 
we assume intra-band transitions dominate this effect, 
then we ignore anomalous velocity, leading $\braket{n\bm{k}|\hat{J}_\alpha|m\bm{k}}\simeq -ev_{\alpha,n\bm{k}}\delta_{m,n}$ where $v_{\alpha,n\bm{k}}\equiv \hbar^{-1}\partial E_{n\bm{k}}/\partial k_{\alpha}$. 
Substituting this in Eq. \eqref{eq:Greenwood-like} and focusing on the dominant band $A$ and $B$, we obtain a simplified formula:
\begin{equation}\label{eq:simplified}
    \sigma_{yx}^{L_\mu}\simeq \frac{e^2\hbar^2}{\pi\Omega}\sum_{\bm{k}} \sum_{n=A,B} \braket{n\bm{k}|\hat{L}_\mu|n\bm{k}}v_{x,n\bm{k}}v_{y,n\bm{k}} \rho_{n\bm{k}}^2,
\end{equation}
indicating that the single band OAM is important. 
It means that this T-odd them requires the SOI to activate the quenched OAM, different to the T-even term in Eq. \eqref{eq:TKNN-like} that does not but relies on inter-band transitions induced by an external field, instead. 
It is also contrary to the MSHE in $\ruo$, that can be obtained by replacing $\hat{L}_\mu$ to $\hat{S}_\mu$ in Eq. \eqref{eq:simplified}, giving a large finite value without SOI. 

\begin{figure}
    \centering
    \includegraphics[width=0.5\linewidth]{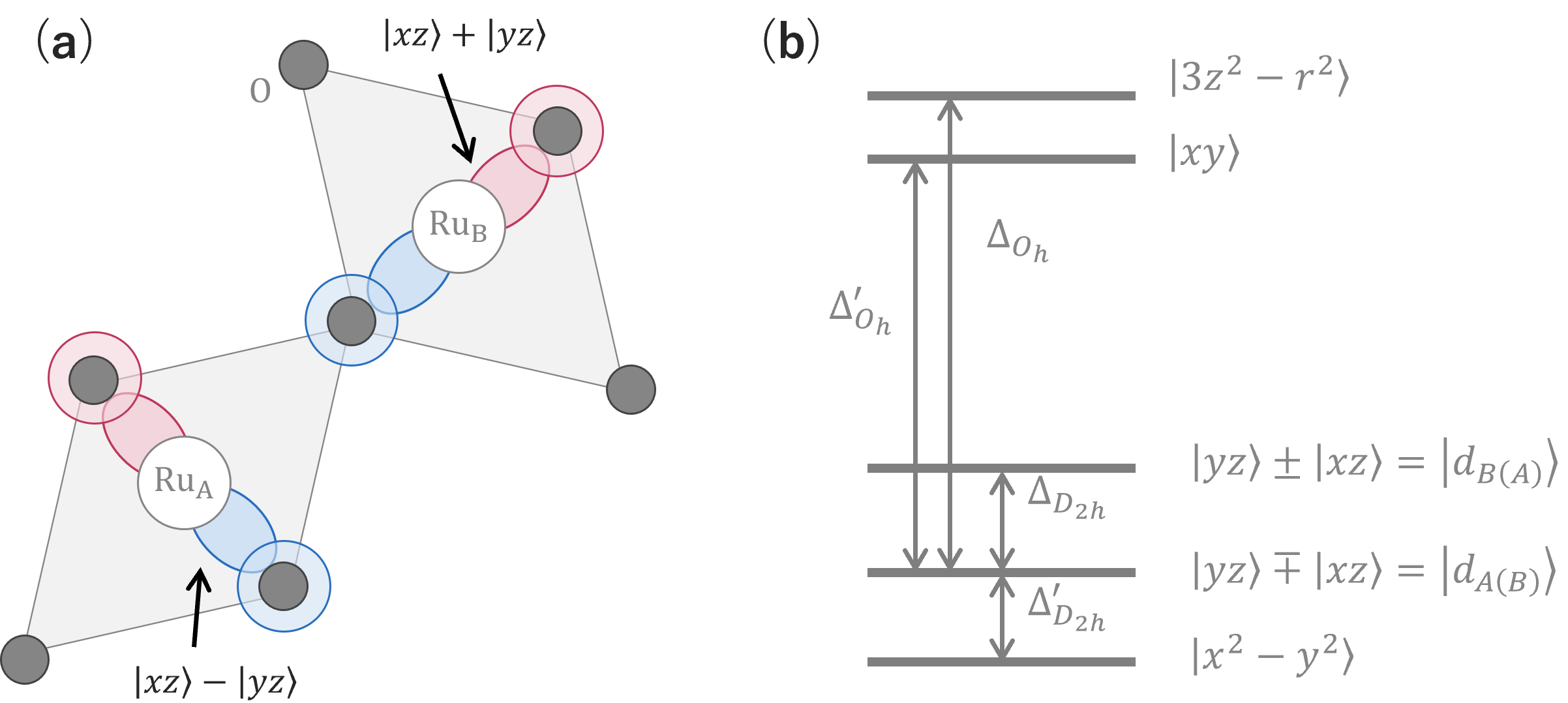}
    \caption{Schematics of the local electronic structure of $\chem{Ru}$-atoms in $\ruo$. (a) $pi$-bonding orders between $\chem{Ru-4d}$ and $\chem{O-2p}$ orbitals, and (b) splitting of $\chem{Ru-4d}$ orbitals under the orthorhombic ($D_{2h}$) ligand fields. }
    \label{fig:ligands}
\end{figure}

To evaluate Eq. \eqref{eq:simplified}, we extract a minimal set of orbitals that consist with the eigen states $\ket{n\bm{k}}$. 
According to a previous study, the bands near the Fermi level are dominated by the $\pi$-bonding between $\chem{O}\haihun2p$ states and the $\chem{Ru}\haihun4d_{t_{2g}}$, especially $d_{zx}$ and $d_{zy}$ states \cite{Berlijn2017-by}.
Reflecting the bond order illustrated in Fig. \ref{fig:ligands} (a), $\ruA$ mainly has $d_{zx}-d_{yz}\equiv d_{A}$ character and $\ruB$ does $d_{zx}+d_{yz}\equiv d_{B}$.
Hence, the eigen states $\ket{n\bm{k}}$ can be well-represented by local atomic orbitals as
\begin{equation} \label{eq:localorbitals}
    \ket{A\bm{k}} \simeq c_{A\bm{k}}^{\ruA,d_{A},\downarrow}\ket{\ruA,d_{A},\downarrow},
    ~~~
    \ket{B\bm{k}} \simeq c_{B\bm{k}}^{\ruB,d_{B},\uparrow}\ket{\ruB,d_{B},\uparrow},
\end{equation}
where $\ket{\ruX,d_{X},\sigma}\ (X=A,B)$ are the atomic orbital basis. 
Note that, the $d_{x^2-y^2}$ character can also be mixed in the vicinity of the $M$-points, but it is only a limited region and is ignored for simplicity.

Let us now consider a perturbation expansion on $\ket{\ruX,d_{X},\sigma}$ with respect to the spin-orbit coupling. 
While the d-orbital splitting arises from both the SOI and the orthorhombic ($D_{2h}$) ligand field, a previous research of the RIXS measurement has clarified that the later is predominant and the SOI can be treated perturbatively \cite{Occhialini2021-ee}. 
This fact allows us setting the $D_{2h}$ splitting states, shown in Fig. \ref{fig:ligands} (b), to the unperturbed states.
Let the SOI Hamiltonian be $H'=\soc\hat{\bm{L}}\cdot\hat{\bm{S}}=\soc(\hat{L}_{x'}\hat{S}_{x'}+\hat{L}_{z'}\hat{S}_{z'}+\hat{L}_{z'}\hat{S}_{z'})$, 
 where $x'$, $y'$, and $z'$ denote the axes of the rotational frame such that $z'$-axis is parallel to $\bm{n}$. 
Considering the perturbation of $H'$ up to the first order, we obtain
\begin{equation} \label{eq:perturbation}
\braket{m\bm{k}|\hat{L}_\mu|m\bm{k}}=-\Re\sum_{d\neq d_m}s_{m}\frac{\soc}{\Delta_d^m} 
\subscripts{0}{\braket{d_m|\hat{L}_\mu|d}}{0}
\subscripts{0}{\braket{d|\bm{n}\cdot\hat{\bm{L}}|d_m}}{0}
+O(\soc^2).
\end{equation}
$\ket{d}_0$ is an unperturbed state where its species $\ruX$ and spin $\sigma$ are abbreviated. 
$s_m$ denotes the spin of the $m$ state with $s_A=-1$ and $s_B=+1$. 
$\Delta_d^m$ is the energy difference between $\ket{d_m}_0$ and $\ket{d}_0$ given in Fig. \ref{fig:ligands} (b). 

Eq. \eqref{eq:perturbation} gives $O(\soc)$ contributions only when $\bm{n}\cdot \hat{\bm{L}}=\hat{L}_\mu$, that is, the orbital polarization is parallel to the Neel vector. 
In addition, as the expectation values of $\hat{L}_z$ and $\hat{L_x}$ are no longer equivalent under the $D_{2h}$ ligand field, Eq. \eqref{eq:perturbation} leads to 
\begin{equation}
    \braket{m\bm{k}|\hat{L}_z|m\bm{k}}\simeq s_m \frac{\soc}{\Delta_{D_{2h}}}, 
    \quad \braket{m\bm{k}|\hat{L}_x|m\bm{k}} \simeq 
    s_m \frac{\soc}{2}\left(\frac{1}{\Delta'_{D_{2h}}}+\frac{1}{\Delta'_{O_{h}}}+\frac{1}{\Delta_{O_{h}}}\right),
\end{equation}
resulting the anisotropy of the OHC as seen in Figs \ref{fig:OHCvsCond}. 
More generally, $\braket{m\bm{k}|\hat{L}_{\mu}|m\bm{k}}\simeq s_m \soc/\Delta_\mu$ where $\Delta_\mu$ is an effective ligand field splitting. 
While the sign of OHC is opposite to that of SHC in Fig. \ref{fig:OHCvsCond}, it is not universal and should depend on the local electronic structure.

It is also apparent that a large MSHE is crucial to obtain a sizable OHE with avoiding cancellation due to $s_m$.
In fact, Eq. \eqref{eq:simplified} may be further approximated to $\sigma_{yx}^{L_\mu} \simeq (\soc/\Delta_\mu)\sigma_{yx}^{S_\mu}$, that is, the MOHE is capped by the MSHE. 
This relation is totally opposite to that between conventional SHE and OHE in which the OHE is more fundamental and induces the SHE \cite{Kontani2009-pn}. 

Together these results provide important insights into the mechanism of this phenomenon as a confluence of the spin splitter effect and the $D_{2h}$ site symmetry through the SOI; 
The large spin current polarizing along the Neel vector direction gathers the parallelly polarized orbital current through the SOI. 
Simultaneously, the quenched OAM is activated anisotropically with respect to the ligand fields.

\paragraph*{Conclusion:}
We have evaluated the orbital Hall effect in the altermagnetic $\ruo$ from first principles based on a combined framework of the density functional theory and the linear response theory. 
The results presented a huge magnetic (time-reversal odd) orbital Hall conductivity, as 10 times larger than those in typical ferromagnetic metals, only when the Neel vector is perpendicular to the $c$-axis, however, it was suppressed when parallel to the $c$-axis, suggesting a strong Neel vector dependence.

Performing a perturbation expansion with respect to the spin-orbit coupling up to the first order, we have identified a mechanism of this behavior as an interplay between the spin splitter effect and an anistropy of orbital angular momentum due to the $D_{2h}$ ligand field. 
In other words, the spin orbit interaction converts a spin current into a parallely polarized orbital current anistropically reflecting a low site symmetry of the $\chem{Ru}$ sites.

From application perspective, these results indicate the importance of choosing the free layer ferromagnet such that it converts the orbital current in the same direction as the spin current. 
When using $\ruo$ as a spin source, a ferromagnet having negative orbit-to-spin conversion, such as the $\chem{Gd}$, is favorable for enhancing its spin torques \cite{Go2020-rb}; Otherwise, a spacer performing the similar role in conversion should be used, inserted \cite{Lee2021-jk}. 
It is also noteworthy that the $\ruo$ or an other d-wave altermagnetic metal can realize an orbital current source offering both controllability and high-efficiency, helping further progress of orbitronics.

\section*{Acknowledgement}
This work was supported by JST-CREST (JPMJCR20C1). The computational results were obtained using HPC resources at Cyberscience Center, Tohoku University.

\bibliography{202409_OHEinRuO2}

\end{document}